\documentclass[11pt,twoside]{article}

\usepackage{asp2014}

\aspSuppressVolSlug
\resetcounters

\begin{document}

\title{Exploring Coronal Heating Using Unsupervised Machine-Learning}

% Note the position of the comma between the author name and the 
% affiliation number.
% Authors surnames should come after first names or initials, eg John Smith, or J. Smith.
% Author names should be separated by commas.
% The final author should be preceded by "and".
% Affiliations should not be repeated across multiple \affil commands. If several
% authors share an affiliation this should be in a single \affil which can then
% be referenced for several author names. If only one affiliation, no footnotes are needed.
% See ManuscriptInstructions.pdf and ASP's manual2010.pdf 3.1.4 for more details

\author{Shabbir Bawaji,$^1$ Ujjaini Alam,$^1$ Surajit Mondal,$^2$ and Divya Oberoi$^2$}
\affil{$^1$Thoughtworks India, Pune, MH, India; \email{shabbirb@thoughtworks.com}, \email{ujjaini.alam@thoughtworks.com}}
\affil{$^2$National Centre for Radio Astrophysics - Tata Institute of Fundamental Research, Pune, MH, India; \email{surajit@ncra.tifr.res.in}, \email{div@ncra.tifr.res.in}}

% This section is for ADS Processing.  There must be one line per author. paperauthor has 9 arguments.
\paperauthor{Shabbir Bawaji}{shabbirb@thoughtworks.com}{}{Thoughtworks India}{Engineering 4 Research (E4R)}{Pune}{Maharashtra}{411006}{India}
\paperauthor{Ujjaini Alam}{ujjaini.alam@thoughtworks.com}{}{Engineering 4 Research (E4R)}{Author2 Department}{Pune}{Maharashtra}{411006}{India}
\paperauthor{Surajit Mondal}{surajit@ncra.tifr.res.in}{}{NCRA-TIFR}{}{Pune}{Maharashtra}{411007}{India}
\paperauthor{Divya Oberoi}{div@ncra.tifr.res.in}{}{NCRA-TIFR}{}{Pune}{Maharashtra}{411007}{India}

% There should be one \aindex line (commented out) for each author. These are used to
% build up the author index for the Proceedings. The surname must come first, followed by
% initials. Note the use of ~ before each initial to control spacing.
% The \author entries (see above) have surname last. These \aindex entries have
% surname first.
% The Aindex.py command willl create them for you after you have constructed the \author
% The first entry should be the first author, for bold-facing the author index page-reference

%\aindex{Bawaji,~Shabbir,~B.~S.}
%\aindex{Alam.~Ujjaini,~A.~U.}
%\aindex{Mondal,~Surajit,~M.~S}
%\aindex{Oberoi,~Divya,~O.~D}

\begin{abstract}
The perplexing mystery of what maintains the solar coronal temperature at about a million K, while the visible disc of the Sun is only at 5800 K, has been a long standing problem in solar physics. 
A recent study by \citet{mondal2020} has provided the first evidence for the presence of numerous ubiquitous impulsive emissions at low radio frequencies from the quiet sun regions, which could hold the key to solving this mystery.
These features occur at rates of about five hundred events per minute, and their strength is only a few percent of the background steady emission.
One of the next steps for exploring the feasibility of this resolution to the coronal heating problem is to understand the morphology of these emissions.
To meet this objective we have developed a technique based on an unsupervised machine learning approach for characterising the morphology of these impulsive emissions.
Here we present the results of application of this technique to over 8000 images spanning 70 minutes of data in which about 34,500 features could robustly be characterised as 2D elliptical Gaussians.
\end{abstract}

% These lines are subject index entries. At this stage these have to commented
% out, and need to be on separate lines. Eventually, they will be automatically uncommented
% and used to generate entries in the Subject Index at the end of the Proceedings volume.
%\ssindex{astronomy!radio!interferometry}
%\ssindex{astronomy!solar!coronal heating}
%\ssindex{techniques!Gaussian fitting}
%\ssindex{methods!statistical!unsupervised clustering}
%\ssindex{algorithm!DBSCAN}

% These lines show examples of ASCL index entries. At this stage these have to commented
% out, and need to be on separate lines. Eventually, they will be automatically uncommented
% and used to generate entries in the ASCL Index at the end of the Proceedings volume.
% The ascl.py command will scan your paper on possible code names.

\section{Introduction}
The solar atmosphere, known as the corona, is at a temperature of about a million K, while the visible disc of the Sun, the photosphere, is only at 5800 K. 
What maintains the coronal temperature so much hotter than the photosphere still remains a puzzle and has come to be known as the {\em coronal heating problem}.
A promising resolution to the problem is now known as the {\em nanoflare hypothesis}, which proposes the presence of a large number of tiny flares taking place all over the Sun all the time, too weak to be detected individually but collectively with sufficient energy to heat up the corona \citep{parker1988}.
Generations of ever more sensitive X-ray and extreme-UV telescopes have however so far been unable to detect these nanoflares.
A key consequence of this hypothesis is the ubiquitous presence of very weak impulsive emissions in the metrewave radio band. 
The very first detection of these emissions has recently been reported by \citet{mondal2020}. 
We refer to these emissions as Weak Impulsive Narrow-band Quiet Sun Emissions (WINQSEs).  
As a step towards exploring in greater detail the suitability of WINQSEs for explaining coronal heating, here we attempt to characterise their morphology.
They have been detected with sufficient SNR in these high fidelity and high time resolution radio images to allow a robust characterisation.
Given the vast volumes of data and the large number of WINQSEs which need to be examined, one necessarily requires a robust automated approach.
Here we present a machine learning based algorithm which meets these requirements and characterises the morphologies of WINQSEs using a 2D elliptical Gaussian model.

\section{Data}
These data span 70 minutes and come from the Murchison Widefield Array \citep{tingay2013} during a period of low solar activity.
They were imaged with a time and frequency resolution of 0.5 s and 160 kHz, leading to over 8000 images at 132 MHz. 
The lone active region present on the visible disc of the Sun appears as the brightest radio source.
Some of the images were completely dominated by the emission from this region.
Such images and the emission from this region are not used for this study. 

\section{Methodology}

The input for our pipeline is a spatio-temporal solar data cube $I(x,y,t)$, where $I$ denotes the flux density, $x$ and $y$ are the spatial coordinates in the image and $t$ the time. 
We start by empirically defining a solar boundary within which to search for peaks. We define a mean image, $P$, over time ($P_{x, y} = \frac{1}{T}\sum_{t=1}^{T} I_{x,y,t}$) and the region in $P$ with intensity $> 0.002 \times P_{max}$ is defined to be the Sun. The same region is used in all the images as the solar boundary.
To avoid working with image features observed with poor Signal to Noise Ratio (SNR), we impose a constraint of SNR $>5$, where SNR is determined from the RMS fluctuations in the map at a location far from the Sun.

On an average, each image has about 10 weak features. 
Naturally some of them tend to be clustered and overlap in the images,  making it harder to model the morphology of individual WINQSEs. 
For ease of morphological characterization, we focus here on the isolated peaks.
To do this efficiently, we use Density-Based Spatial Clustering of Applications with Noise (DBSCAN), an unsupervised, non-parametric clustering algorithm \citep{ester1996density}. 
Given a set of points in some parameter space, this algorithm groups together points with many nearby neighbours, and isolated points lying in regions of low-density are treated as outliers. 
Here, DBSCAN is applied such that if at least two peaks are less than $5$ pixels apart they are regarded as a cluster, and the peaks that are farther apart are marked as isolated ones by the algorithm.

The next step is to find an optimal window size for fitting Gaussians to the individual peaks. 
A window that is too small would not have enough information for a successful fitting, while a large window could potentially include contaminating emission from neighboring peaks. During peak finding, the identified peaks are characterized by multiple parameters -- peak amplitude, peak importance (difference between the peak and the average surrounding flux), distance to the next nearest peak, peak SNR, etc.
We transform this many-parameter system to a two-parameter system using t-distributed Stochastic Neighbor Embedding or tSNE \citep{vanDerMaaten2008}. 
This algorithm is a nonlinear dimensionality reduction technique that models each higher dimensional object by a two-dimensional point. It calculates the probabilities of similarity between points in the higher and lower dimensional spaces, then minimizes the difference between these probabilities. This leads to an accurate two-dimensional representation of the original data. The lower dimensional data is now clustered using DBSCAN and the optimal fitting window-size of each cluster is determined to be the median nearest-neighbour-distance for that cluster.

We find that a 2D elliptical Gaussian, defined below, provides a good description for the vast majority of isolated features.
\begin{equation}
    I = O + A \frac{2}{\pi^2 \sigma_x \sigma_y} e^{\left[-\frac{1}{2} \left(\frac{1}{\sigma_x} [(x-x_0) \cos \theta  - (y-y_0) \sin \theta]\right)^2 \right. \nonumber\\
\left. -\frac{1}{2} \left(\frac{1}{\sigma_y} [(y-y_0) \cos \theta  + (x-x_0) \sin \theta]\right)^2  \right]},
\end{equation}
where $O$ is the offset, $A$ the amplitude, $\sigma_x, \sigma_y$ the widths of the major and minor axes, $x_0, y_0$ the position of the Gaussian peak, and $\theta$ its angle of orientation with respect to the x-axis.

\begin{figure}%
    \centering
    {{\includegraphics[width=6.5cm]{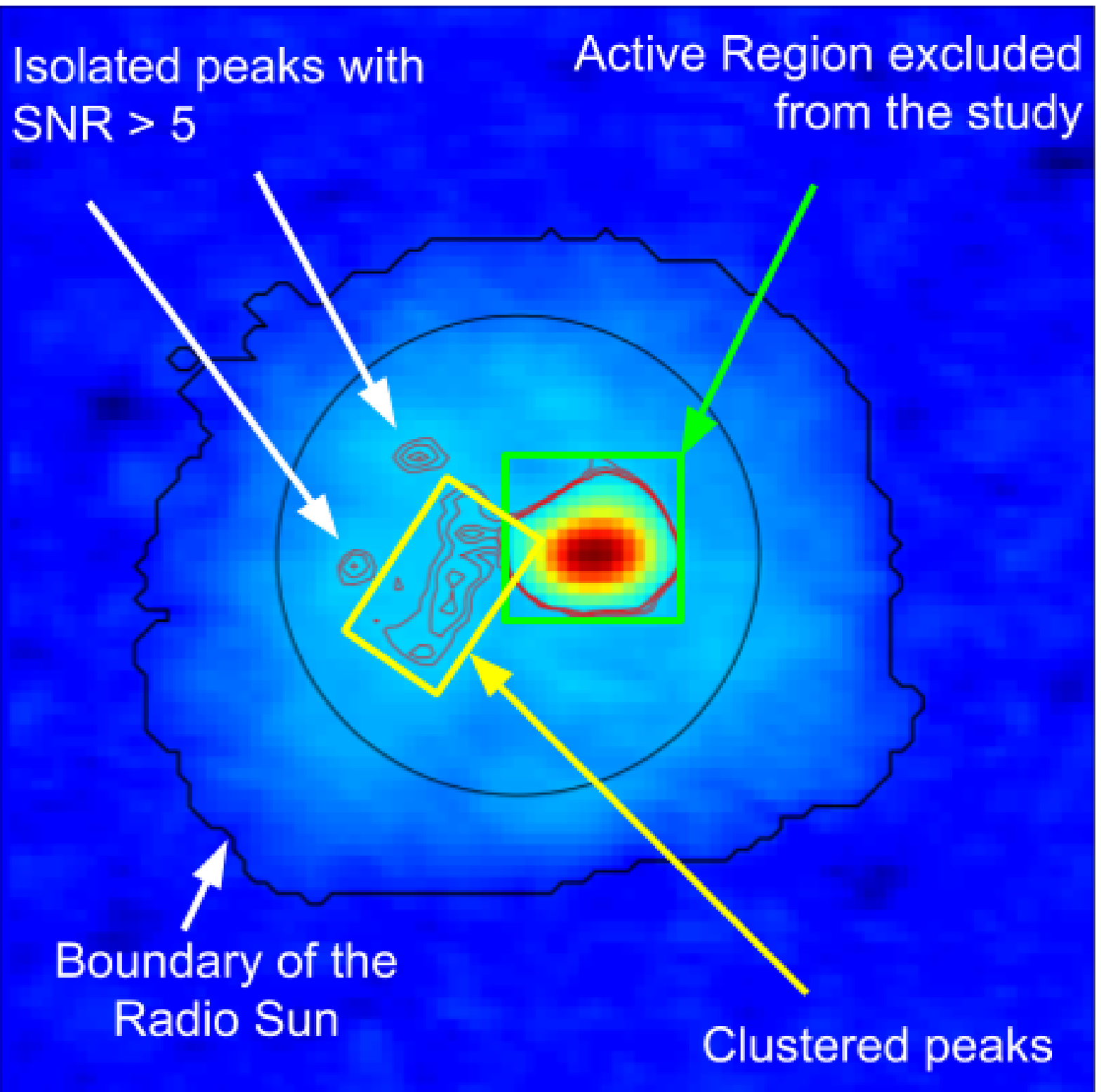} }}%
    \qquad
    {{\includegraphics[width=4.5cm]{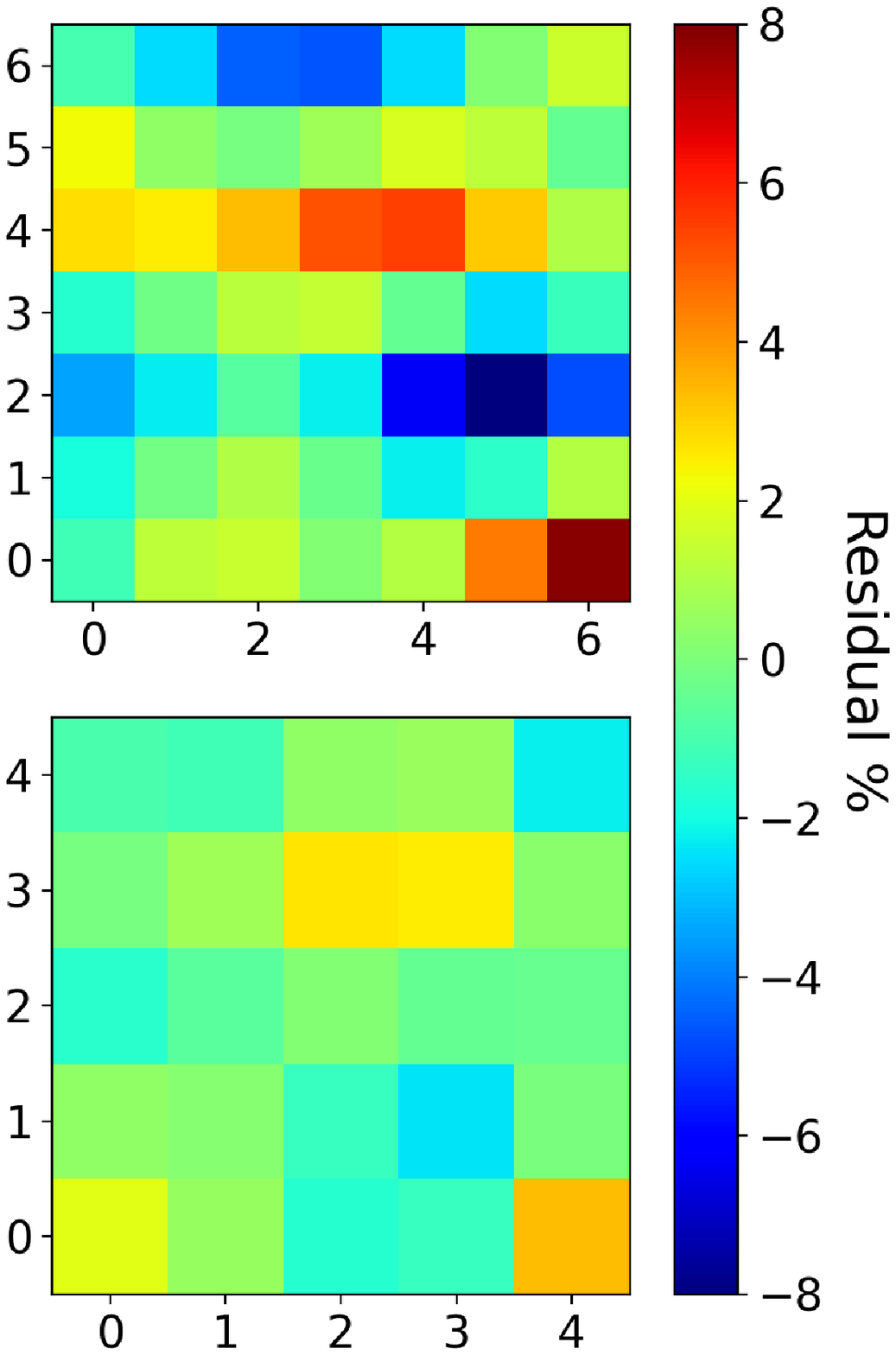} }}%
    \caption{The left panel shows a typical radio image of the Sun. Solar boundary in visible and radio parts of the spectrum are marked. Examples of both isolated and clustered peaks are indicated. The right panel shows the residuals for the two isolated peaks fit by the 2D elliptical Gaussian model.}
    \label{fig:example}
\end{figure}

The algorithm further rejects outliers with large errors on the Gaussian fit parameters, or with a large $\chi^2$, and outputs a set of Gaussian parameters that can be used to characterize WINSQEs. 
An example radio image is shown in Fig.~\ref{fig:example} along with the residuals for the best fit Gaussian fits to the two isolated features.
The residuals are only at a few percent, confirming that the peaks are described well by the Gaussian models.

\section{Results}

\begin{figure}%
    \centering
    {{\includegraphics[width=4.55cm]{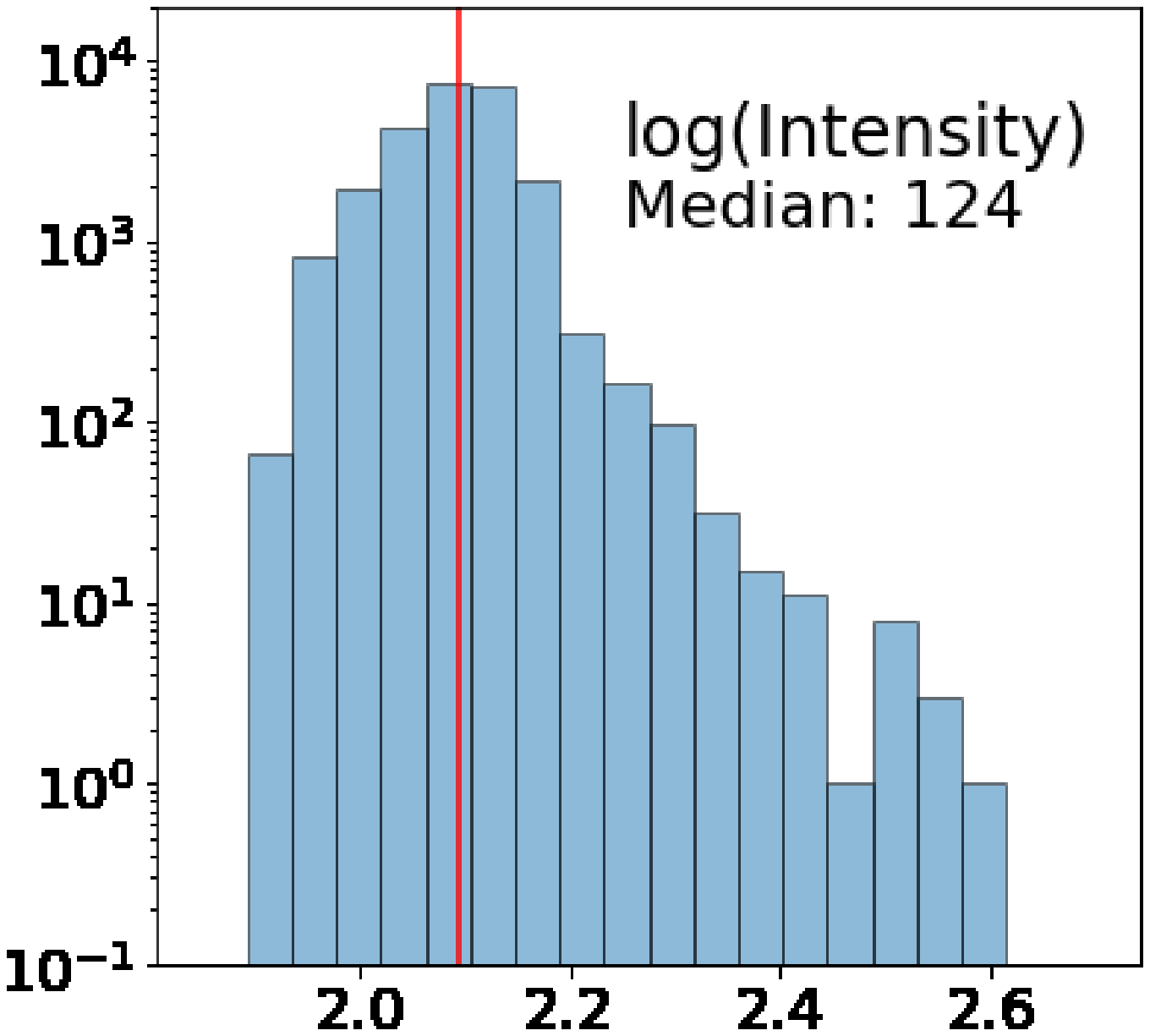} }}%
    {{\includegraphics[width=4.55cm]{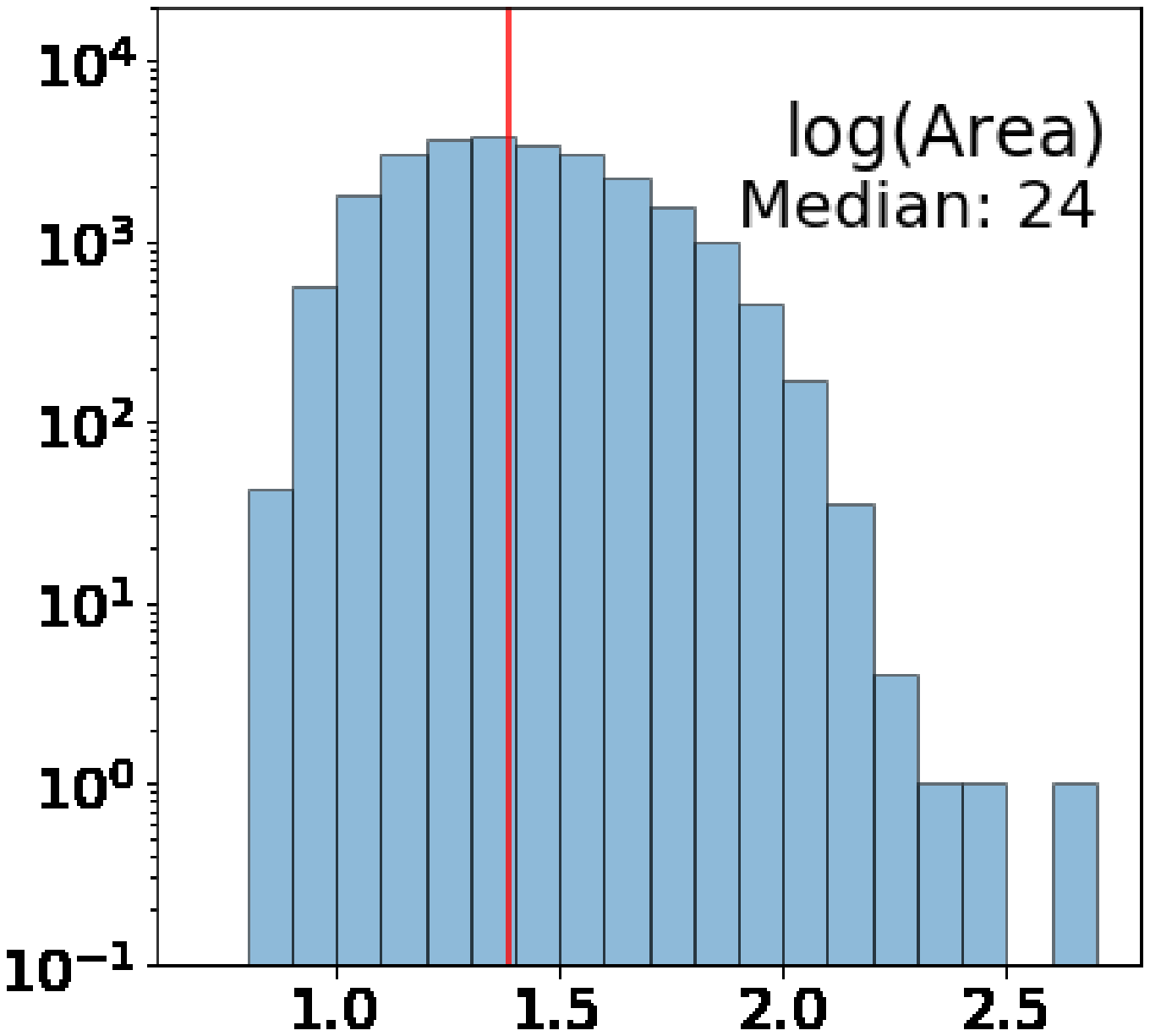} }}\\
    {{\includegraphics[width=4.55cm]{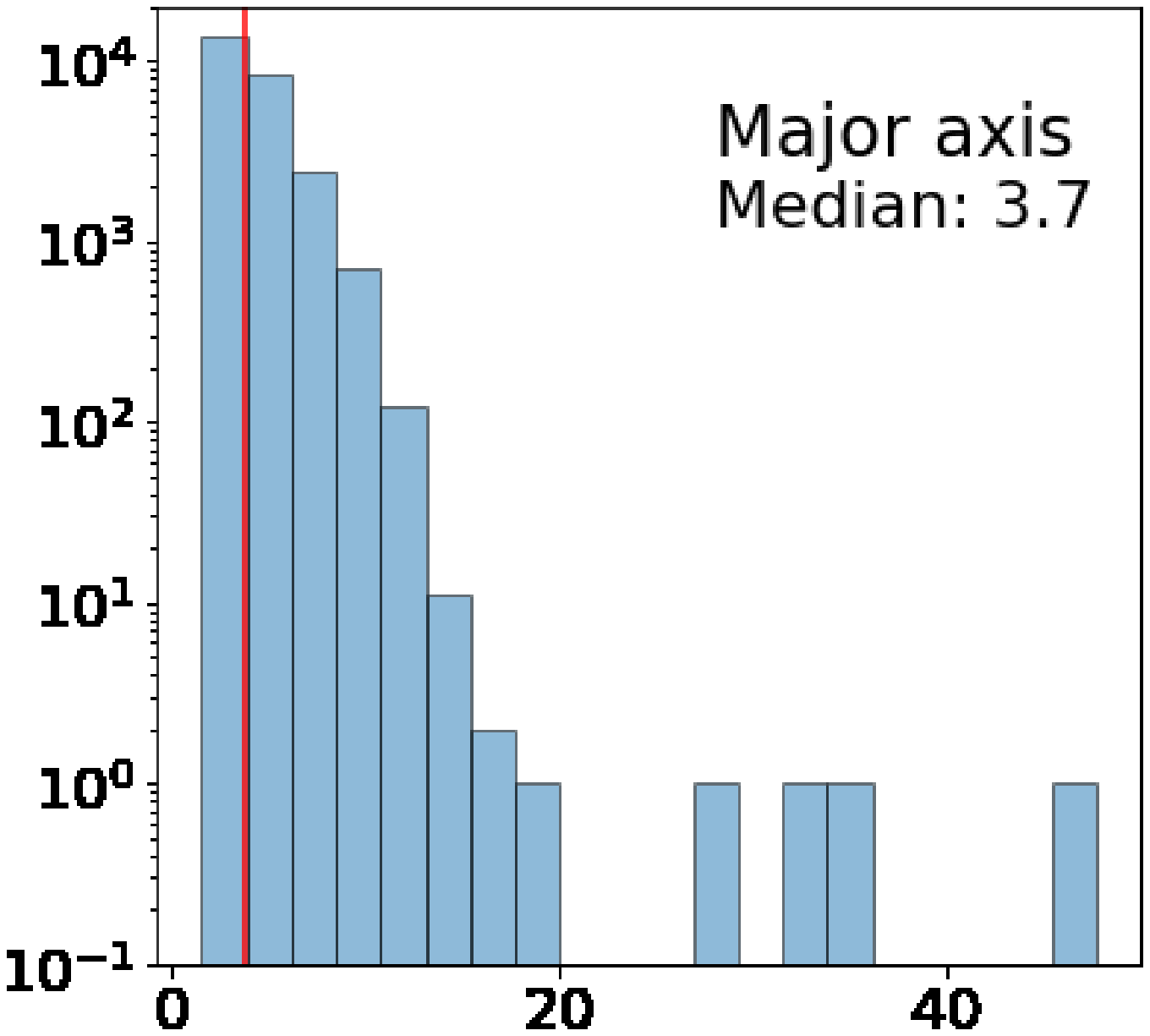} }}
    {{\includegraphics[width=4.55cm]{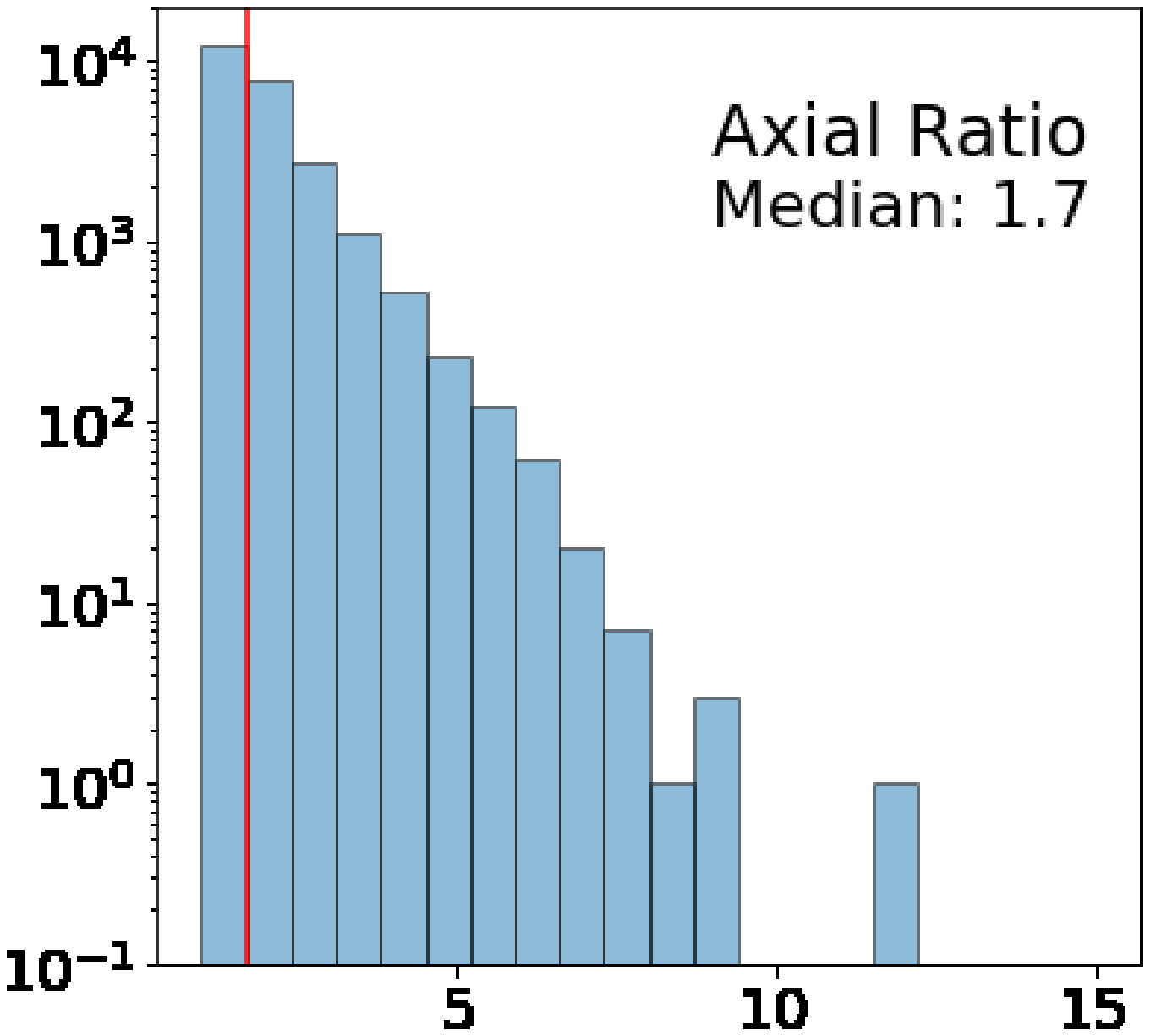} }}
    \caption{Distributions of the best fit Gaussian parameters for the WINSQEs (intensity, area, major axis and axial ratio). 
    The median values are mentioned in linear units and marked by a vertical red line in each of the panels.
    }
    \label{fig:hist}
\end{figure}

A total of 70,845 peaks were found with $SNR > 5$. 
Of these, 42,469 were identified to be isolated. 
The $\chi^2$ for Gaussian fits to 34,457 of these peaks were found to be acceptably low. Figure~\ref{fig:hist} shows the distributions of intensity, area, axial ratio and major axis of the best fit Gaussian parameters. 
The size and areas of the Gaussian models 
show that their emissions are quite compact, with median area of $\sim24$ and median major axis width of $\sim4$ in pixel units, while the angular resolution of these observations is $\sim14$ in the same units.
This is in line with theoretical expectations.  Each of these distributions show a small number of outliers with unphysically large values, these are under investigation.

\section{Conclusions}

The algorithm presented here successfully distinguishes between the weak isolated and clustered emissions and provides a robust characterisation of the shapes of isolated WINQSEs. 
The modeled characteristics are found to be in line with the theoretical expectations for the origin of WINQSEs.
Our immediate next step will be to test the robustness of the pipeline with additional data-sets.
From a science perspective, it will be very interesting to study the evolution of morphology of these sources across both time and frequency axes, and the MWA is capable of providing data suitable for such an exploration.
We plan to expand the scope of this work to meet this objective.

% \section{Conference Photographs}

% At the end of this template you may find a commented line with the bookpartphoto where the editors could decided to add a conference photo,
% might there be enough room at the end of your paper. Leave this comment in, do not supply your own photos in this paper but contact the editors
% if you have some interesting shots.

\acknowledgements 
SB acknowledges H. Hayatnagarkar, S. Khandekar, D. Singh and S. Surana, all from Thoughtworks India, for useful discussions.

\bibliography{P5-83.bib} 

% if we have space left, we might add a conference photograph here. Leave commented for now.
% \bookpartphoto[width=1.0\textwidth]{foobar.eps}{FooBar Photo (Photo: Any Photographer)}

\end{document}